\documentstyle{elsart}
\begin{document}
\input psfig.sty

\tolerance=500
\hoffset=0pt
\voffset=0pt
\hyphenation{cal-o-rim-e-ter}
\hyphenation{Methyl-cou-ma-rin}
\hyphenation{Di-phe-nyl-ox-a-zole}
\hyphenation{mon-o-en-er-get-ic}
\hyphenation{pos-i-tron}
\begin{frontmatter}

\title{Light response of pure CsI calorimeter crystals \\
painted with wavelength-shifting lacquer}

\author[UVa]{E. Frle\v z\thanksref{author}},
\thanks[author]{Corresponding author; Tel: +1--804--924--6786, 
fax: +1--804--924--4576, e--mail: frlez@virginia. edu (E. Frle\v{z})\hfill}
\author[PSI]{Ch.~Br\"onnimann},
\author[PSI]{B.~Krause},
\author[UVa]{D.~Po\v cani\'c},
\author[PSI]{D.~Renker},
\author[UVa,PSI]{S.~Ritt},
\author[UVa,PSI]{P.~L.~Slocum},
\author[IRB]{I.~Supek},
\author[PSI]{H.~P.~Wirtz}
\address[UVa]{Department of Physics, University of Virginia, 
Charlottesville, VA~22904, USA}
\address[PSI]{Paul Scherrer Institute, Villigen PSI, CH-5232, Switzerland}
\address[IRB]{Institute Rudjer Bo\v skovi\'c, Bijeni\v cka 46, 
HR-10000 Zagreb, Croatia}

\begin{abstract}
We have measured scintillation properties of pure CsI crystals used in 
the shower calorimeter built for a precise determination of the 
$\pi^+$$\rightarrow$$\pi^0e^+\nu_e $ decay rate at the Paul Scherrer Institute
(PSI). All 240 individual crystals painted with a special wavelength-shifting 
solution were examined in a custom-build detection apparatus (RASTA---radioactive 
source tomography apparatus) that uses a $^{137}$Cs radioactive gamma 
source, cosmic muons and a light emitting diode as complementary 
probes of the scintillator light response. We have extracted the total light 
output, axial light collection nonuniformities and timing 
responses of the individual CsI crystals. These results predict 
improved performance of the 3$\pi\,$sr PIBETA calorimeter due to 
the painted lateral surfaces of 240 CsI crystals. 
The wavelength-shifting paint treatment did not affect appreciably 
the total light output and timing resolution of our crystal sample. 
The predicted energy resolution for positrons and photons in the energy 
range of 10--100$\,$MeV was nevertheless improved due to the more favorable 
axial light collection probability variation. We have compared simulated 
calorimeter ADC spectra due to 70$\,$MeV positrons and photons with a Monte Carlo 
calculation of an ideal detector light response. 
\par\noindent\hbox{\ }\par\noindent
\par\noindent\hbox{\ }\par\noindent
\end{abstract}
\end{frontmatter}
\vfill\eject

\section{Introduction}

The PIBETA detector at the Paul Scherrer Institute in Switzerland
was designed for the precise measurement of the pion beta ($\pi\beta$) decay 
branching ratio ($\pi^+$$\rightarrow$$\pi^0e^+\nu_e$) as well as for study of 
other rare pion and muon decays~\cite{Poc88}. The experiment has started 
to take production data in the summer of 1999. The heart of the detector 
is the $\sim$3$\pi\,$sr pure cesium iodide (CsI) segmented calorimeter. 
In-depth knowledge of the calorimeter response to the positrons and photons 
in the energy range 10--100$\,$MeV is essential in order 
to achieve the proposed accuracy of the branching ratio measurements
of $\sim$0.5$\,$\%.
Experimental signature of a $\pi\beta$ event is determined by the prompt
decay $\pi^0 \to \gamma\gamma$. The (1.025$\pm\,$0.034)$\times$$10^{-8}$ pion 
beta decay branching ratio~\cite{McF85} will be remeasured relative to 
the $10^{4}$ times more probable $\pi^+$$\rightarrow$$e^+\nu_e$\/ decay 
rate, that is known with the combined statistical and systematic 
uncertainty of $\sim\,$0.33$\,$\%~\cite{Cza93,Bri94}. 
 
The detector energy response functions were determined both experimentally, by 
scanning the calorimeter with the monoenergetic $e^+$ and $\gamma$ beams, and 
in Monte Carlo simulations. Inputs to the simulation calculations 
were the average light outputs and light output nonuniformity functions 
for each individual calorimeter module. In this paper we describe how these 
input parameters, essential for a realistic Monte Carlo detector simulation,
have been measured in a dedicated tomography apparatus. Section~\ref{sec:surf}
details the surface treatment of the CsI scintillator crystals. 
Design and operation of the radioactive source scanning apparatus 
are explained in Sec.~\ref{sec:rasta}. The extracted light output parameters are
discussed in Secs.~\ref{sec:light} and \ref{sec:nonunif}. In Sec.~\ref{sec:geant} we 
explain how these parameters are used in a {\tt GEANT}-based calculation of 
the calorimeter response.

\section{CsI Crystal Surface Treatment}\label{sec:surf}

The PIBETA calorimeter consists of 240 truncated crystal pyramids with a height
of 22$\,$cm, which corresponds to 12 radiation lengths. Nine different 
detector shapes were required to complete the crystal sphere~\cite{Ken76}: four 
irregular hexagonal truncated pyramids (we labeled them HEX--A, HEX--B, HEX--C, 
and HEX--D), one regular pentagonal (PENT) and two irregular half-hexagonal 
truncated pyramids (HEX--D1 and HEX--D2), and two trapezohedrons (VET--1, VET--2). 
The volumes of studied CsI crystals vary from 797$\,$cm$^3$ (HEX--D1/2) 
to 1718$\,$cm$^3$ (HEX--C).

Twentyfive pure CsI crystals were produced by Bicron Corporation facility in 
Newbury, Ohio. The reminder of the crystals were grown and cut in the Institute for 
Single Crystals (AMCRYS) in Harkov, Ukraine. The more detailed description of 
the PIBETA calorimeter design is given in Refs.~\cite{Ass95}~and~\cite{Frl99}. In 
Fig.~\ref{fig:ball} we show the partially assembled calorimeter with front 
faces and wrapped side surfaces of exposed CsI modules visible inside 
the mechanical support structure.

The light output of each crystal, namely the average number 
of photoelectrons for unit energy deposition caused by ionizing charged
particles, can be tuned by using the different polishing, matting and wrapping 
treatments of the detector surfaces. The goal of such treatments is usually to maximize 
the light yield thus obtaining the best energy resolution, while keeping 
the detector response linear irrespective of the location of the energy deposition. 

Our initial treatment of the crystal surfaces involved a test sample of 74 CsI 
detectors and consisted of: (a) polishing the crystal surfaces with a mixture 
of 0.2 $\mu$m aluminum oxide powder and etylenglycol, (b) wrapping the lateral 
surfaces with two layers of Teflon foil plus one layer of aluminized Mylar 
sheet, and (c) covering the front crystal surface with a black paper 
template. The crystal back surface with the glued phototube was left uncovered.
We reported the predicted response of the calorimeter with all CsI modules treated 
in the described manner in our previous paper~\cite{Frl99}.

Further test measurements and the technical developments~\cite{ren95} suggested
an improved optical treatment of crystal surfaces. Due to the fact that CsI  
is slightly hygroscopic, it is desirable to ``seal'' the scintillators 
against potential surface deterioration and thus minimize 
changes of the detector response throughout the duration of the experiment.
Traditional polishing and wrapping methods cannot ensure long term stability 
of the relevant scintillation parameters due to the insufficient chemical resistance of
alkali halides~\cite{Bri93}. Furthermore, a suitable protective coating of the crystal surfaces
can at the same time be used to modulate the scintillator light collection 
properties and can be easily removed, if necessary.  

We therefore decided to paint the surfaces of every CsI crystal using 
a special organosilicon mixture. The solution was developed  
at the Harkov Single Crystals Research Institute~\cite{xxx}. For our purposes the most 
useful polymers were polyphenylsiloxanes due to their high thermal stability, good radiation 
resistance and the moisture-proof quality. The optical properties of the organosilicon coatings 
depend on the nature of the organic radicals. Their phenyl polymer bonds can 
efficiently transfer the energy to the luminescent admixtures~\cite{gry94}.
The selected polymers and solvents have a nonpolar character resulting in a low moisture 
absorption. On the other hand, the high optical transparency of the varnishes was achieved 
in a purification process by the absorption trapping of silica, aluminum oxide and hydroxide 
impurities. The resulting absorption and reemission of light by the coating in the wavelength range 
above 300$\,$nm was better than 99\%.

Specifically, the waveshift lacquer was the ladder organosilicon copolymer with the 
chemical composition PPO+POPOP+COUM.1, where PPO is 2.5--Di\-phe\-nyl\-ox\-a\-zole, POPOP 
represents 1.4-Di-2-(5-Phe\-nyl\-ox\-a\-zolile-Benzene) and COUM.1 is 
7-Diethylamine-4-Methylcoumarin~\cite{hand71}. That paint solution was applied twice 
resulting in a protective cover with a total thickness of 120 $\mu$m. The dried paint 
layer has the index of refraction $n_D$=1.62 and the density $\rho$=1.094, and is
soluble in ethyl alcohol. Following the paint application each detector module 
was wrapped in two layers of a 38$\,$$\mu$m PTFE Teflon sheet and two layers of 
20$\,\mu$m thick aluminized Mylar to make it optically insulated from its 
calorimeter neighbors.

We have compared the average fast-to-total (F/T) analog signal ratio and 
the photoelectron statistics of all lacquer-treated crystals with the measurements
of polished and Teflon-wrapped crystals we reported on in Ref.~\cite{Frl99}.
The F/T ratio was virtually unchanged at 0.77, while the
number of photoelectrons per MeV was just under 3$\,$\% lower, changing
from an average of 64 to 62 Photoelectrons/MeV at the ambient temperature of 
18$^\circ$C. The combined statistical and systematic uncertainty of these 
measurements was less then 1 Photoelectron/MeV.

The averaged lineshape of an analog cosmic muon signal from one CsI detector
captured on a digital oscilloscope is shown in Fig.~\ref{fig:waveform}.
The 10-to-90$\,$\% amplitude rise time of the analog signal changed from 4.0$\,$ns
for the Teflon-wrapped scintillators to 4.8$\,$ns for the lacquer-painted detectors,
due to the fact that a fraction of the scintillator light is now absorbed 
by the wavelength-shifting surfaces and reemitted later with 
the different wavelength. Histograms of the F/T ratios and the photoelectron 
numbers for the painted crystals are shown in Figs.~\ref{fig:ft}~and~\ref{fig:np},
respectively.

\section{RASTA: RAdioactive Source Tomography Apparatus}\label{sec:rasta}

The total light output and the position-dependent light collection probabilities for 
each CsI detector were measured in a test setup using a $^{137}$Cs radioactive 
gamma source with an activity of 3.7 MBq. It was found that this method produced results 
consistent with the cosmic muon tomography~\cite{Frl99} and was well suited for the routine 
crystal uniformity tests. Thus an apparatus was designed and built to allow 
a fast, precise and reproducible method of determining the optical nonuniformity of 
each CsI crystal. 

The drawing of the main parts of the apparatus is shown in Fig.~\ref{fig:apparatus}.
The $^{137}$Cs radioactive gamma source was embedded in a lead collimator. The collimator 
had a thickness of 5$\,$cm while its pin-hole had a diameter of 6$\,$mm. The probability for 
the 0.662$\,$MeV $^{137}$Cs photons to penetrate that lead thickness is $<1\,$\%. 
The ionization energy deposited at the central axis of a CsI crystal is 
about one-sixth of the energy deposition near the crystal surface.
The collimator was mounted on an aluminum plate, which was moved by a stepping motor. 
The precision of the step motor movements was ~12.5$\,\mu$m. The examined crystal was placed on 
a horizontal aluminum plate using a dedicated positioning system with a precision of 
$\sim$0.3$\,$mm. The stepping motor was controlled by a PC486 personal computer through
a RS232 interface. 

The schematic diagram of the logic electronics is shown in Fig.~\ref{fig:electronics}. 
The analog photomultiplier signal was amplified by a factor of 10 using an LRS 612A
photomultiplier amplifier. The signal was then split, one copy going into the trigger logic 
while the other copy was delayed and fed into an LRS 2249A CAMAC ADC. The trigger created 
a computer LAM in the SIN IO506 input/output unit. The CAMAC modules were controlled using a
HYTEC 1331 Turbo CAMAC--PC interface.  

The stability of the system after switching on the PMT high voltage was
studied in preliminary calibrations. The data were taken with a CsI crystal 
in a fixed position every 5 minutes and the results are shown in
Fig.~\ref{fig:hv}. We concluded that the PMT high voltage should be switched on for 
at least one hour before doing any measurements with the detector. 

A full calibration was performed on each CsI crystal by scanning it four times within 
three hours. 
The measured light output as a function of scan position are shown in Fig.~\ref{fig:reproduc}
for the CsI detector number {\tt S040}. After the cross-normalization the results of different 
measurements agreed within the 0.5$\,$\% statistical uncertainty.

\section{Total Light Output of the CsI Detectors}\label{sec:light}

Scintillation light of a full-sized CsI crystal was read by a 78$\,$mm
diameter EMI 9821QKB photomultiplier tube with quartz photocathode
window. The smaller half-sized crystals were directly coupled to a 52$\,$mm $\phi$ 
EMI 9211QKA PMTs with similar characteristics~\cite{pmt}. Both phototube types were 
glued to the back faces of the crystals using a clear silicone elastomer 
Sylgard 184 manufactured by Dow Corning~\cite{dow}. 

A light-emitting diode (LED) was attached to the exposed back side of
the CsI detector. The whole detector assembly was placed in the light-tight RASTA box.
The LEDs were excited at a $\sim$10$\,$Hz rate using a NIM diode driver
with an adjustable output voltage. The LED pulses simulated constant
energy depositions in the CsI crystal corresponding to a 10--100$\,$MeV range.
The absolute energy calibration was established by comparison with measured cosmic muon
ADC spectra. These spectra were acquired in calibration runs triggered by two
small plastic scintillator tag counters placed above and below the CsI crystal. Their
expected peak positions, expressed in the units of MeV, were calculated in
a {\tt GEANT} Monte Carlo simulation.

For LED measurements the variance $\sigma_E^2$\/ of a PMT pulse-height spectrum depends 
upon the mean number of photoelectrons $\bar N_{pe}$\/ from the photocathode created per
unit energy deposition in the scintillator:
\begin{equation}
\sigma_E^2=\sum_i\sigma_i^2+\bar E/\bar N_{pe},
\label{eq:led}
\end{equation}
where $\bar E$\/ is the spectrum peak position and $\sigma_i^2$'s are 
assorted variances, such as the instabilities of LED driving voltage and 
temporal ADC pedestal variations.

The energy resolutions $\sigma_E$'s were measured for 9 different values of the LED 
driving voltage and the photoelectron numbers $\bar N_{pe}$ were calculated by the
least-squares fit using Eq.~(\ref{eq:led}).
The mean numbers of photoelectrons normalized to the unit equivalent-energy deposition
in MeV and scaled down the ambient temperature of 18$^\circ$C are shown 
in Fig.~\ref{fig:np}. The histograms for the full-sized crystals and half-sized 
crystals are displayed separately, with the ratio of 2.4 in photoelectron statistics 
essentially consistent with the surface area ratio of 2.25 in the two differently-sized 
photocathodes. 

\section{Spatial Light Output Nonuniformity of the CsI Detectors}\label{sec:nonunif}

The 3-dimensional spatial distribution of scintillation light output 
can be specified by giving the number of photoelectrons $N_{\rm pe}(x,y,z)$ produced 
by a 1$\,$MeV energy deposition at the point ($x,y,z$) (``3-dimensional light 
nonuniformity function''). 
In the following discussion we limit ourselves to the linear and one-dimensional variation 
of the detected light output $N_{\rm pe}(z)$ along the long axis of the detector:
\begin{eqnarray}
N_{\rm pe}(z)\propto\cases{
N_1+a_{z1}\cdot z&  $0\le z\le 10\ {\rm cm}$, \cr
N_2+a_{z2}\cdot z& $10\le z\le 18\ {\rm cm}$, \cr
N_3+a_{z3}\cdot z& $18\le z\le 22\ {\rm cm}$  \cr
}
\end{eqnarray}
where $a_z$'s are the linear optical nonuniformity coefficients in
the axial coordinate $z$, and the coordinate system origin is at
the center of the detector front face. In this approximation the transverse
light output variations are integrated away. In our figures and tables we
are quoting the $a_z$ light nonuniformity parameters in units of \%/cm.

The complete RASTA database of all 240 studied CsI crystals is made available for
online inspection at the PIBETA WWW site~\cite{data}. The measurements
for each CsI crystal are summarized on separate WWW pages.
Individual pages start with the pulse-height spectra recorded
with the cosmic muon trigger that is used for the absolute energy calibration
of the ADC scale. The LED ``energy'' spectra corresponding to 9 different LED intensities
are shown next, together with the superimposed Gaussian fits. The least-square linear
fits of the data using the Eq.~(1), which allow the deduction of the average number of 
photoelectrons per MeV, as well as the associated fit uncertainties, are documented below 
the spectra. The timing resolution histogram is also presented; it gives the standard 
deviation of the CsI detector timing spectrum with respect to the tag counters TDC 
value for the cosmic muon triggers.

Also displayed are the ADC spectra taken with the $^{137}$Cs radioactive source 
at 11 equidistant positions along the long axis of a studied crystal, together
with Gaussian-exponential fits. The fitted Gaussian peak positions 
are plotted in the axial light nonuniformity figures that follow next.
The relative axial variations of the light collection nonuniformities
are shown for 6 representative CsI crystals in Fig.~\ref{fig:nonunif}.
The extracted linear nonuniformity coefficients $a_z$'s are listed last.

The relation between the linear nonuniformity coefficients $a_{z1}$ in the front
part of the crystal ($z\le10\,$cm) and the coefficients $a_{z2}$ describing the
position-dependent change of the axial light output in the central and back
crystal part is shown on the scatter plot in Fig.~\ref{fig:a1a2} for all 240 
crystals. Fig.~\ref{fig:a1a2_fun} compares the linear function 
$a_{z2}=c_1+c_2\cdot a_{z1}$ fitted to the scatter plot with the result deduced 
from our previous measurements of the Teflon-wrapped crystals, Ref.~\cite{Frl99}.
The parameters $c_1$ and $c_2$ were determined by the ``robust'' estimation
method~\cite{Hub81} by imposing the requirement of the minimum absolute deviation 
between the measured and calculated values. The treatment of crystal 
surfaces with the wavelength-shifting lacquer produces more favorable light output 
collection along the detector axis by increasing the collected light from the
back part of a crystal and thus compensating for the energy leakage of 
electromagnetic showers and improving the detector energy resolution. These results were
confirmed in the full-fledged {\tt GEANT} simulation described in Sec~\ref{sec:geant}.

In Fig.~\ref{fig:time} we show the timing resolutions for the all 
measured  CsI crystals. We have histogrammed the standard deviations of the Gaussian functions
fitted to the spectrum of the timing differences between the CsI detector and the small
scintillator tag counter just above the crystal. The average CsI detector timing uncertainty specified in
such a way is 0.68$\,$ns.

We have summarized the average scintillation properties of all 240 studied CsI crystals in 
Tables~\ref{tab1}~and~\ref{tab2}.

\section{Simulated ADC Lineshapes of 70 MeV/c $e^\pm$'s and $\gamma$'s}\label{sec:geant}

We have developed a comprehensive {\tt GEANT} Monte Carlo description of the PIBETA 
detector that includes all major active as well as passive detector components~\cite{Bru94}. 
The user code is written in a modular form in standard {\tt FORTRAN} 
and organized into over 300 subroutines and data files~\cite{Frl97}. The graphical
user interface to the code based on {\tt Tcl} routines supplements the program. 
  
Standard {\tt GEANT} routines account for electromagnetic shower processes in CsI: leak-through, 
lateral spreading, backsplash, as well as the electromagnetic interactions in the active target and 
tracking detectors: bremsstrahlung, Bhabha scattering, in-flight annihilation for positrons, 
and pair production, Compton scattering, etc. for photons.

The individual PIBETA detector components, both passive and active ones, can be positioned or 
``switched off'' without recompiling the code. The following detectors are defined in 
the Monte Carlo geometry:
\begin{enumerate}
\item a beam counter, an active plastic scintillator degrader and several versions 
of segmented stopping active targets;
\item two concentric cylindrical multiwire proportional chambers used for charged particle
tracking;
\item a 20-piece cylindrical plastic scintillator hodoscope used for charged particle 
detection and discrimination;
\item a 240-module pure CsI calorimeter sphere;
\item a 5-plate cosmic muon veto scintillator system;
\item passive calorimeter support structure and the individual detector support systems, 
\item the detector phototubes and HV divider bases, and the lead brick shielding structure.
\end{enumerate}

User input to the simulation code requires the detector version, e.g., 
the run year, the beam properties, the selected reaction/decay final states, and the 
version of the ADC and TDC simulation codes. User can select one particular final state 
or any combination of different final states with the relative probabilities defined 
by the reaction cross sections and  decay branching ratios. The simulation of ADC values
and TDC hits takes into account the individual detector photoelectron statistics, the axial 
and transverse light collection detector nonuniformities, ADC pedestal variations, 
electronics noise, and the event pile-up effects. The photoelectron statistics and
the individual detector light collection nonuniformity coefficients are initialized
from the RASTA database file. Calculation of accidental coincidences up to 
the fourth order is optional.

Selectable options also include the effects of:
\begin{enumerate}
\item lateral and axial extent and the divergences of the stopping pion beam;
\item positron and muon beam contamination; 
\item photonuclear and electron knockout reactions in CsI material~\cite{photo};
\item aluminized Mylar wrapping of CsI modules and plastic scintillator veto staves;
\item cracks between the CsI detector modules;
\item temperature coefficients of the individual calorimeter modules;
\item gain instability of the calorimeter detector modules; 
\item electronics discriminator thresholds and ADC gate width. 
\end{enumerate}

Particular attention was paid to the correct accounting of ``software gains'' for
the individual CsI modules. In the simulation calculation the values of the detector 
software gains allow the same user control element as the detector high voltages in
the operation of the physical detector. The software gains are determined in an iterative
procedure constrained by the Monte Carlo $\pi\to e^+\nu$ positron ADC spectra in
each CsI detector. Namely, in the real experiment the high voltages of the
individual CsI detectors are also set by matching the positions of the 69.8 MeV positron peaks. 

The output of the calculation provides the selected experimental layout and the
final states chosen in the calculation, as well as the used production cross sections
and branching ratios. A number of physical variables that could be of interest to a user
are histogrammed in 1- and 2-dimensional format. Raw-wise and column-wise 
{\tt PAW} Ntuples~\cite{Bru93} are used to digitize the individual events. Simulated 
energy depositions and the ADC and TDC values associated with the active detectors 
are saved on an event-per-event basis. A single event display program can be used to 
examine in detail the particular event in an output stream.

Simulated calorimeter ADC spectra for 69.8$\,$MeV monoenergetic positrons and 70.8$\,$ MeV 
photons generated in the stopping target are shown in Fig.~\ref{fig:tails1}. 
The figure shows the ADC lineshapes summed over all 220 calorimeter signals that make up
the low threshold ($E_{\rm THR}$$>$$5\,$MeV) trigger logic. The top panel shows the spectra without 
the energy cut on the 20 summed veto CsI signals that define the beam opening/exit calorimeter 
surfaces, the bottom panel shows the same spectra with the imposed veto energy cut. 
Fig.~\ref{fig:tails2} displays the same variables but now summed only over the calorimeter detector 
with the maximum energy deposition and its nearest neighbors. That restricted ADC sum should
reduce the contribution of accidental event pile-up to the smearing of the energy spectra. 

The $e^+$ and $\gamma$ peak positions, their widths and tail fractions of the simulated ADC 
lineshapes are summarized in Tables~\ref{tab3}~and~\ref{tab4}. We see that the calorimeter energy 
resolution is improved and the energy spectra tail fractions reduced in the detector
with the paint-treated crystals. The improvement is smaller for the positron showers which
deposit most of their energy close to the incident particle impact point, but is considerable,
over 15$\,$\%, for the 70$\,$MeV photons which penetrate deeper into the calorimeter.

\section{Conclusion}

We have measured the scintillation properties of the individual pure CsI
crystals used in the PIBETA calorimeter. The crystals were treated with a special
wavelength-shifting organosilicon paint. The fraction of the fast decay component
in the total light output, the photoelectron statistics and the temperature dependence 
of the light output are documented for the each detector. A $^{137}$Cs radioactive source
scanning apparatus was designed and used to determine the axial light collection nonuniformity
of all scintillator crystals. An individual CsI detector response was described by three linear 
light nonuniformity coefficients.

These experimentally determined scintillation properties of the calorimeter components
provide the essential input to a comprehensive {\tt GEANT} Monte Carlo of the PIBETA detector. 
We have compared the simulated performance of the PIBETA calorimeter having 
the waveshift-painted CsI crystals with an idealized calorimeter characterized by 
the uniform light collection from every volume element of the scintillator.
The comparison shows that the PIBETA calorimeter made up of the painted CsI crystals 
has better energy resolution, while its timing resolution is not compromised.

\section{Acknowledgements}

This work is supported and made possible by grants from the US National
Science Foundation and the Paul Scherrer Institute.
\bigskip

\section{Appendix A. RASTA Programs}

The main computer program which operated the apparatus was named RASTA. Its main tasks were to: 
\begin{enumerate}
\item control the stepping motor operation;
\item control and read out the CAMAC modules; 
\item perform histogram analysis including pedestal and background subtraction,
      histogram filling and fitting;
\item provide graphical data display. 
\end{enumerate}

Upon the execution of the program, the operator was first asked for the serial number {\tt Sxxx} of
the crystal being tomographed. A new disk subdirectory {\tt hisxxx} was created in which 
all the relevant histograms of a scan process were stored. 

The starting parameters of the program were defined in the file {\tt RASTA.INI}:
{\tt
\begin{verbatim}
[Scan]
Time=20
Origin=1.0
Stepsize=2.0
NumberofSteps=11
[Fit]
PeakFindMin=10.
PeakFindMax=800.
Transition=-20.0
NsigmaXLow=2.0
NsigmaXHigh=6.0
\end{verbatim}
}

The category {\tt Scan} controlled the scanning process: 
\begin{enumerate}
\item {\tt Time} is the scan time per point in seconds; 
\item {\tt Origin} is the starting point of the scan (with the origin marked 
      in the drawing of the apparatus, Fig.~\ref{fig:apparatus}); 
\item {\tt Stepsize} is the distance between measured points given in centimeters.
\end{enumerate}

The category {\tt Fit} executed the fitting process: 
\begin{enumerate}
\item {\tt PeakFindMin} ({\tt PeakFindMax}) is the lower (upper) limit of the peak position
      passed to the fitting program when searching for the maximum ADC value and FWHM 
      of the pulse-height spectrum;
\item {\tt Transition} is the value of the Gauss--exponential transition of 
      the fit function measured from the peak position; 
\item {\tt NsigmaXLow} and {\tt NsigmaXHigh} define the histogram range 
      to be fitted in units of the Gaussian standard deviation from the peak position. 
\end{enumerate}

After having changed or accepted the default run parameters the step motor was 
initialized and the collimator moved to the parking position where 
the pedestal data and background data were taken. 

The crystal was then scanned automatically at the points specified in the {\tt RASTA.INI}
file. Fig.~\ref{fig:tomodis3} shows the computer display as it appeared during the scanning
process.

At the end of the scan all histograms were stored so that the data could be easily reviewed 
in a Microsoft {\tt EXCEL} spreadsheet (histograms numbers 0 and 1) or with the in-house 
developed Histogram Manager utility~\cite{ritt95} (histograms 2 to NSTEP+3). 

Two other programs are supplied together with the {\tt RASTA} package: 
\begin{enumerate}
\item {\tt HVSET}, which helps select the high voltage (HV) value for the photomultiplier 
      attached to a new crystal being prepared for tomography. The additional input parameters 
      are the HV starting value, the preset pulse-height spectrum peak value and maximum differ
ence
      between the preset and fitted peak value; 
\item {\tt MOVE}, which allows to position the collimator at the required 
      location.
\end{enumerate}

\vfill\eject

\clearpage

\vspace*{\stretch{1}}
\begin{figure}[!tpb]
\caption{Picture of the CsI calorimeter during the mechanical assembly. About 
half of individual crystals (of the 240 total) are seen in a self-supporting
configuration inside the detector cradle.}
\label{fig:ball}
\end{figure}

\begin{figure}[!tpb]
\caption{Comparison of the cosmic muon waveforms for a hexagonal
CsI detector wrapped in aluminized Mylar (dashed line) and for the same
crystal painted with the wavelength-shifting lacquer (full line).}
\label{fig:waveform}
\end{figure}

\begin{figure}[!tpb]
\caption{Histogram of the fast-to-total light component ratio for all studied
CsI crystals. Light output of unwrapped painted crystals was measured with 
a digital oscilloscope at the average ambient temperature of 22$^\circ$C.}
\label{fig:ft}
\end{figure}

\begin{figure}[!tpb]
\caption{Histogram of the number of photoelectrons per MeV of deposited energy for full-sized 
(left panel) and half-sized (right panel) CsI detectors normalized to
the ambient temperature of 18$^\circ$C.}
\label{fig:np}
\end{figure}

\begin{figure}[!tpb]
\caption{Schematic drawing of the RASTA experimental apparatus. The depicted
assembly is enclosed in a light-tight aluminum box with a hinged cover plate.}
\label{fig:apparatus}
\end{figure}

\begin{figure}[!tpb]
\caption{Schematic drawing of the RASTA logic electronics. In routine
data acquisition mode the trigger is provided by the discriminated single 
CsI detector signal.}
\label{fig:electronics}
\end{figure}

\begin{figure}[!tpb]
\caption{Plot of the drift and leveling off of the pulse-height spectrum peak
position due to the settling down of the photomultiplier during the first hour 
after the PMT high voltage is switched on.}
\label{fig:hv}
\end{figure}

\begin{figure}[!tpb]
\caption{Plots of the reproducibility of the axial light nonuniformity measurements
demonstrated with four sets of measurements of a single CsI crystal taken
consecutively.}
\label{fig:reproduc}
\end{figure}

\begin{figure}[!tpb]
\caption{Plots of the relative light output as a function of the radioactive source 
axial position for six different CsI detectors. The detector phototubes
are positioned at $z$=22$\,$cm.}
\label{fig:nonunif}
\end{figure}

\begin{figure}[!tpb]
\caption{Light nonuniformity coefficients $a_{z1}$ and $a_{z2}$
(\%/cm), determined separately for the front ($z$$\le$10 cm) and back
($z$$\ge$10 cm) crystal sections, are plotted in a scatter-plot for
240 CsI detectors. Full-sized hexagonal and pentagonal detectors are
represented with full circles, while the open circles indicate
the half-hexagonal and trapezial detector shapes.}
\label{fig:a1a2}
\end{figure}

\begin{figure}[!tpb]
\caption{Plot of the relationship between the linear light output nonuniformity
coefficients $a_{z1}$ and $a_{z2}$ for the polished and Teflon-wrapped CsI 
crystals (dotted line) and for crystals painted with the waveshift lacquer 
(full line).}
\label{fig:a1a2_fun}
\end{figure}
\vspace*{\stretch{2}}
\clearpage

\vspace*{\stretch{1}}
\begin{figure}[!tpb]
\caption{Cosmic muon timing for one representative CsI crystal with respect 
to the plastic scintillator tag counter (top panel) and the timing resolutions
of all studied CsI crystals (bottom panel). The resolution is defined as 
the standard deviation of the TDC timing difference
between the tag counter and the CsI detector.}
\label{fig:time}
\end{figure}

\begin{figure}[!tpb]
\caption{The predicted PIBETA calorimeter spectra for monoenergetic
69.8$\,$MeV positrons and 70.8$\,$MeV photons. All CsI values above the
0.5$\,$MeV TDC threshold (excluding the 20 trapezial veto counters) 
were summed to obtain the deposited energy. 
The {\tt GEANT} simulation used different numbers of photoelectrons/MeV 
and linear axial light collection nonuniformities for each crystal 
as extracted in the RASTA analysis. The bottom panel shows the spectrum 
obtained when the CsI calorimeter veto cut is applied.}
\label{fig:tails1}
\end{figure}

\begin{figure}[!tpb]
\caption{The predicted PIBETA calorimeter spectra for monoenergetic
69.8$\,$MeV positrons and 70.8$\,$MeV photons. ADC values for the CsI crystal
with the maximum ADC value together with ADC readings of its nearest
neighbors were summed. A {\tt GEANT} simulation used different
numbers of photoelectrons/MeV and linear axial light collection
nonuniformities for each crystal, as measured in the RASTA apparatus. 
The dashed histograms represent the simulation with the software detector 
gains equal to 1.}
\label{fig:tails2}
\end{figure}

\begin{figure}[!tpb]
\caption{The RASTA computer screen display showing the fitted histograms during 
automated data acquisition and analysis of a single CsI crystal.}
\label{fig:tomodis3}
\end{figure}

\vspace*{\stretch{2}}
\clearpage

\vspace*{\stretch{1}}
\bigskip
\begin{table}[ph]
\caption{Average scintillation properties of the painted hexagonal 
and pentagonal PIBETA CsI calorimeter crystals PENTAs, HEX--As, 
HEX--Bs, HEX--Cs and HEX--Ds (200 crystals).
The light yields are normalized to the temperature of 18$^\circ$C. 
All other parameters were measured at the average laboratory room
temperature of 22$^\circ$C.}
\label{tab1}

\begin{tabular}{lcc}
\hline
\multicolumn{1}{c}{\ } 
&\multicolumn{1}{c}{BICRON CsI} 
&\multicolumn{1}{c}{Harkov CsI}\\
\multicolumn{1}{c}{\ } 
&\multicolumn{1}{c}{Crystals (22)} 
&\multicolumn{1}{c}{Crystals (178)}\\
\hline\hline
Fast-to-Total Ratio (100 ns/1 $\mu$s gate)   &  0.835    & 0.760   \\
No. photoelectrons/MeV (100 ns ADC gate)      &  74.2     & 65.2    \\
No. photoelectrons/MeV (1 $\mu$s ADC gate)    &  88.8     & 85.8    \\
Fast Light Temp. Coefficient (\%/$^\circ$C)  &  $-1.20$  & $-1.61$ \\
Total Light Temp. Coefficient (\%/$^\circ$C) &  $-1.35$  & $-1.40$ \\
Axial Nonuniformity Coefficient (\%/cm),     &           &         \\
\ \ \ $z$$\le$10 cm, 100 ns ADC gate         &  $0.178$  & $0.297$ \\
Axial Nonuniformity Coefficient(\%/cm),      &           &         \\
\ \ \ $z$$\ge$10 cm, 100 ns ADC gate         &  $0.405$  & $0.635$ \\
\hline
\end{tabular}
\end{table}
\vspace*{\stretch{2}}
\clearpage

\vspace*{\stretch{1}}
\bigskip
\begin{table}[ph]
\caption{Average scintillation properties of the lacquered half-hexagonal 
and trapezial PIBETA CsI calorimeter modules HEX--D1/2s and VETO--1/2s 
(40 crystals). All crystals were polished and painted and wrapped in 
an aluminized Mylar foil afterwards.}
\label{tab2}

\begin{tabular}{lcc}
\hline
\multicolumn{1}{c}{\ } 
&\multicolumn{1}{c}{BICRON CsI} 
&\multicolumn{1}{c}{Harkov CsI}\\
\multicolumn{1}{c}{\ } 
&\multicolumn{1}{c}{Crystals (3)} 
&\multicolumn{1}{c}{Crystals (37)}\\
\hline\hline
Fast-to-Total Ratio (100 ns/1 $\mu$s gate)   &  0.806    & 0.757   \\
No. photoelectrons/MeV (100 ns ADC gate)      &  31.3     & 26.2    \\
No. photoelectrons/MeV (1 $\mu$s ADC gate)    &  38.8     & 34.6    \\
Fast Light Temp. Coefficient (\%/$^\circ$C)  &  $-2.08$  & $-0.63$ \\
Total Light Temp. Coefficient (\%/$^\circ$C) &  $-1.78$  & $-0.69$ \\
Axial Nonuniformity Coefficient (\%/cm),     &           &         \\
\ \ \ $z$$\le$10 cm, 100 ns ADC gate         &  $0.062$  & $0.315$\\
Axial Nonuniformity Coefficient(\%/cm),      &           &         \\
\ \ \ $z$$\ge$10 cm, 100 ns ADC gate         &  $1.592$  & $1.735$ \\
\hline
\end{tabular}
\end{table}
\vspace*{\stretch{2}}
\clearpage

\vspace*{\stretch{1}}
\bigskip
\begin{table}
\caption{The predicted energy resolutions and tail contributions for
69.8 MeV $e^+$ and 70.8 MeV $\gamma$'s events in the full PIBETA
calorimeter. The light output in photoelectrons/MeV and the linear axial
light collection nonuniformities measured for individual CsI crystals 
($a_{z1,z2}$={\tt RASTA}) were used in the {\tt GEANT} simulation. 
Values corresponding to perfect optically homogeneous crystals 
($a_{z1,z2}$=0) are shown for comparison.}

\label{tab3}
\begin{tabular}{lcccc}
\hline
\multicolumn{1}{c}{Parameter} 
&\multicolumn{1}{c}{69.8 MeV $e^+$}
& 
&\multicolumn{1}{c}{70.8 MeV $\gamma$} 
& \\ 

\multicolumn{1}{c}{No $E_V$ cut} 
&\multicolumn{1}{c}{$a_{z1,z2}$=0} 
&\multicolumn{1}{c}{{\tt RASTA} $a_{z1,z2}$} 
&\multicolumn{1}{c}{$a_{z1,z2}$=0} 
&\multicolumn{1}{c}{{\tt RASTA} $a_{z1,z2}$} \\
\hline\hline
Peak Position (MeV)         & 68.87$\pm$0.03& 68.78$\pm$0.03
 & 70.10$\pm$0.03& 69.50$\pm$0.05 \\
FWHM$_{(220)}$ (MeV)        &  3.66$\pm$0.03&  3.50$\pm$0.03
 &  3.98$\pm$0.03&  3.97$\pm$0.04 \\
5$\le$Events$\le$54 MeV (\%)&  6.44$\pm$0.09&  6.11$\pm$0.09
 &  8.20$\pm$0.10&  7.32$\pm$0.09 \\
5$\le$Events$\le$55 MeV (\%)&  6.89$\pm$0.09&  6.48$\pm$0.09
 &  8.86$\pm$0.10&  7.83$\pm$0.10 \\

\hline
\multicolumn{1}{c}{$E_V$$\le$5 MeV} 
&\multicolumn{1}{c}{$a_{z1,z2}$=0} 
&\multicolumn{1}{c}{{\tt RASTA} $a_{z1,z2}$} 
&\multicolumn{1}{c}{$a_{z1,z2}$=0} 
&\multicolumn{1}{c}{{\tt RASTA} $a_{z1,z2}$} \\

Peak Position (MeV)         & 68.88$\pm$0.03& 68.89$\pm$0.03
 & 70.07$\pm$0.03& 69.50$\pm$0.04 \\
FWHM$_{(220)}$ (MeV)        &  3.67$\pm$0.03&  3.49$\pm$0.03
 &  3.99$\pm$0.03&  3.97$\pm$0.04 \\
5$\le$Events$\le$54 MeV (\%)&  1.91$\pm$0.05&  1.70$\pm$0.05
 &  3.92$\pm$0.07&  3.15$\pm$0.07 \\
5$\le$Events$\le$55 MeV (\%)&  2.24$\pm$0.05&  1.99$\pm$0.05
 &  4.46$\pm$0.07&  3.57$\pm$0.07 \\
\hline
\end{tabular}
\end{table}
\vspace*{\stretch{2}}
\clearpage

\vspace*{\stretch{1}}
\bigskip
\begin{table}
\caption{The predicted energy resolutions and tail contributions for
69.8 MeV $e^+$ and 70.8 MeV $\gamma$ events in PIBETA clusters
containing the crystal with maximum energy deposition and its nearest 
neighbors.}
\label{tab4}
\begin{tabular}{lcccc}
\hline
\multicolumn{1}{c}{Parameter} 
&\multicolumn{1}{c}{69.8 MeV $e^+$}
& 
&\multicolumn{1}{c}{70.8 MeV $\gamma$} 
& \\ 

\multicolumn{1}{c}{No $E_V$ cut} 
&\multicolumn{1}{c}{$a_{z1,z2}$=0} 
&\multicolumn{1}{c}{{\tt RASTA} $a_{z1,z2}$} 
&\multicolumn{1}{c}{$a_{z1,z2}$=0} 
&\multicolumn{1}{c}{{\tt RASTA} $a_{z1,z2}$} \\
\hline\hline
Peak Position (MeV)         & 67.74$\pm$0.03& 67.80$\pm$0.04
 & 68.80$\pm$0.03& 68.63$\pm$0.03 \\
FWHM$_{\rm (NN)}$ (MeV)     &  5.22$\pm$0.03&  5.37$\pm$0.04
 &  6.01$\pm$0.03&  5.98$\pm$0.04 \\
5$\le$Events$\le$54 MeV (\%)&  3.83$\pm$0.07&  3.67$\pm$0.07
 &  5.47$\pm$0.08&  4.67$\pm$0.08 \\
5$\le$Events$\le$55 MeV (\%)&  4.54$\pm$0.07&  4.64$\pm$0.08
 &  6.33$\pm$0.09&  5.39$\pm$0.09 \\

\hline
\multicolumn{1}{c}{$E_V$$\le$5 MeV} 
&\multicolumn{1}{c}{$a_{z1,z2}$=0} 
&\multicolumn{1}{c}{{\tt RASTA} $a_{z1,z2}$} 
&\multicolumn{1}{c}{$a_{z1,z2}$=0} 
&\multicolumn{1}{c}{{\tt RASTA} $a_{z1,z2}$} \\

Peak Position (MeV)         & 67.75$\pm$0.03& 67.76$\pm$0.03
 & 68.80$\pm$0.04& 68.60$\pm$0.04 \\
FWHM$_{\rm (NN)}$ (MeV)     &  5.52$\pm$0.03&  5.24$\pm$0.03
 &  5.92$\pm$0.03&  5.85$\pm$0.03 \\
5$\le$Events$\le$54 MeV (\%)&  3.53$\pm$0.07&  3.17$\pm$0.07
 &  5.28$\pm$0.08&  4.29$\pm$0.08 \\
5$\le$Events$\le$55 MeV (\%)&  4.19$\pm$0.07&  3.76$\pm$0.07
 &  6.10$\pm$0.09&  4.98$\pm$0.08 \\
\hline
\end{tabular}
\end{table}
\vspace*{\stretch{2}}
\clearpage

\vspace*{\stretch{1}}
\centerline{\psfig{figure=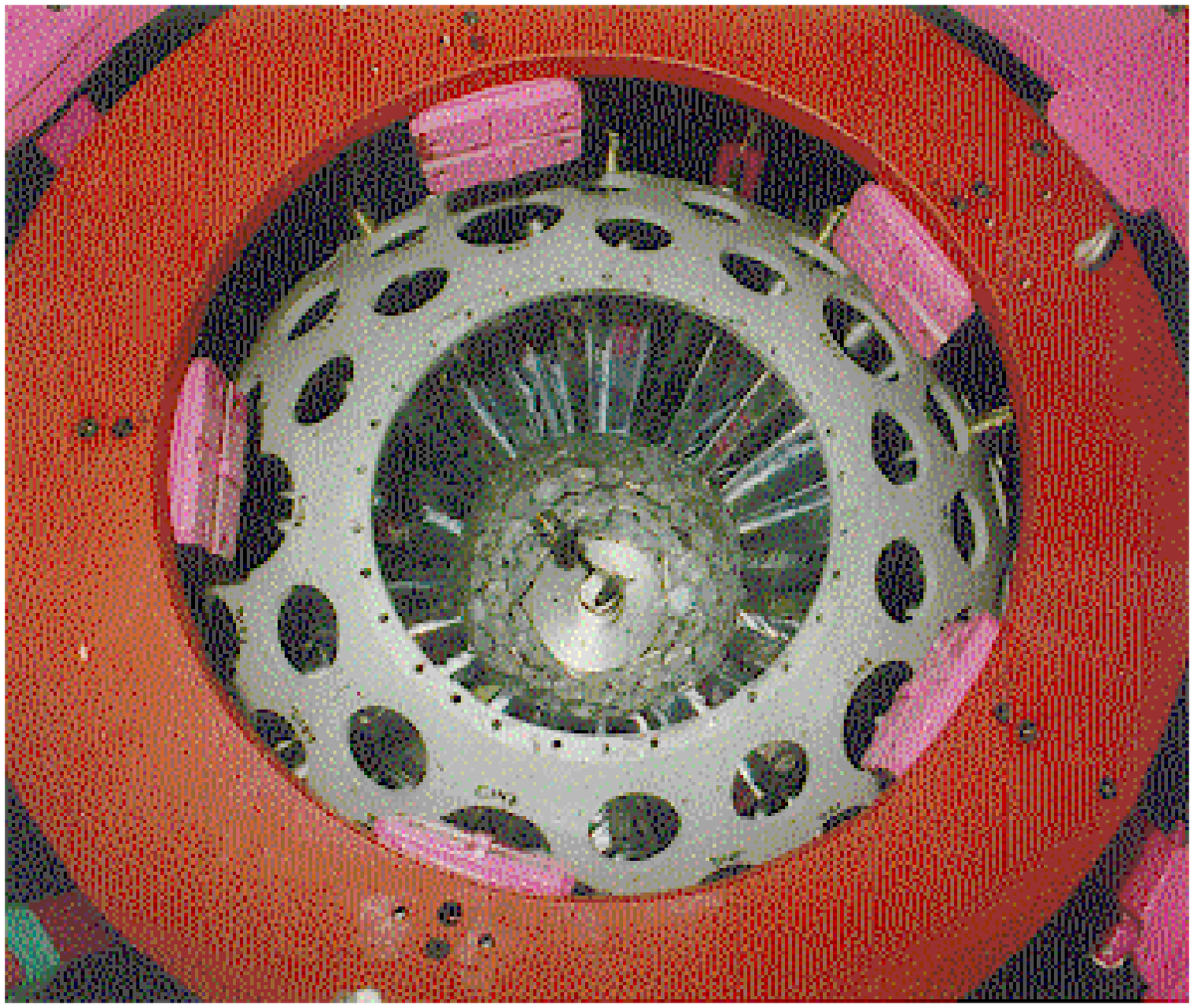,height=10cm}}
\vglue 2.0cm
\centerline{FIGURE 1}
\vspace*{\stretch{2}}
\clearpage

\vspace*{\stretch{1}}
\centerline{\psfig{figure=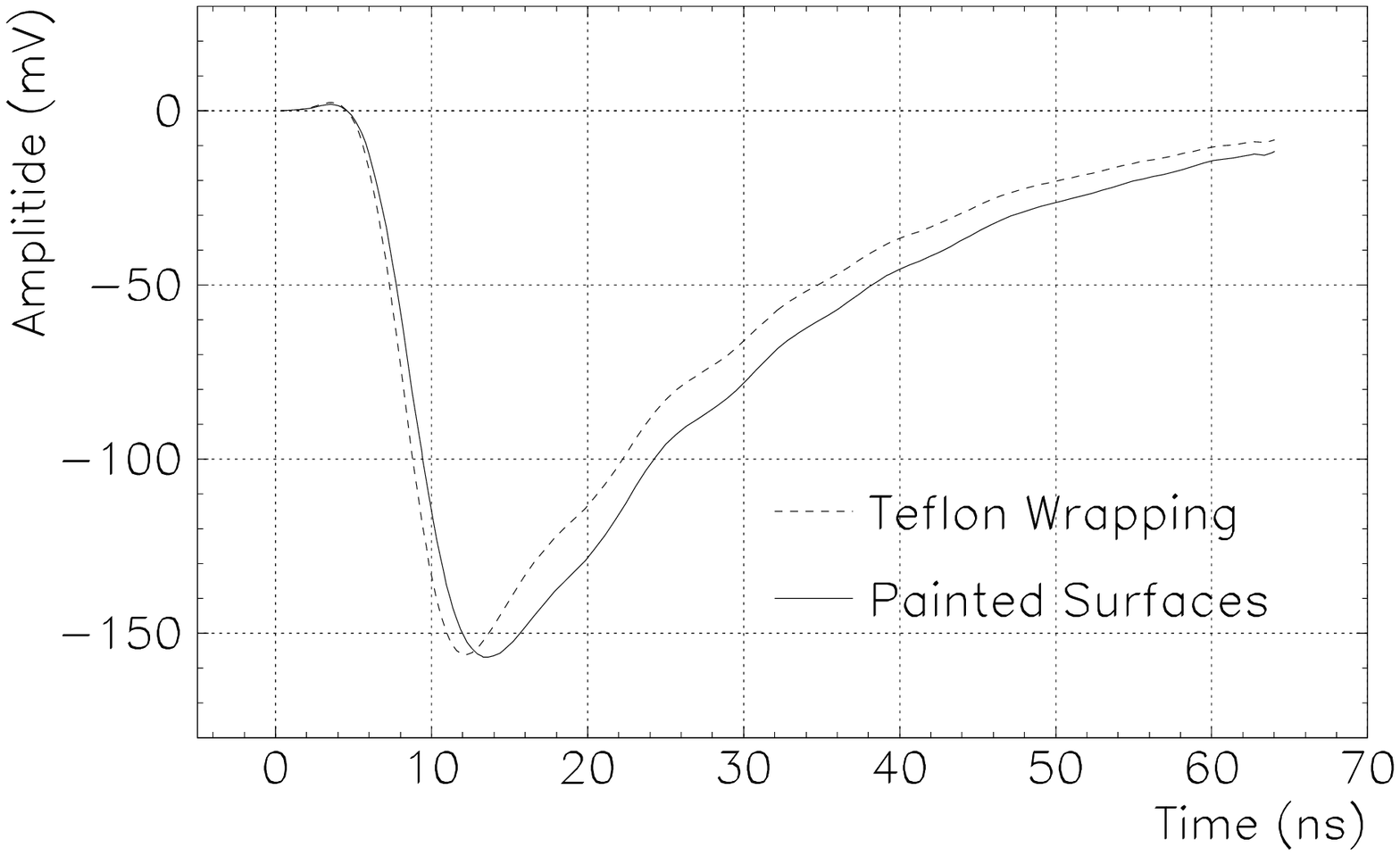,height=20cm}}
\vglue -8.5cm
\centerline{FIGURE 2}
\vspace*{\stretch{2}}
\clearpage

\vspace*{\stretch{1}}
\centerline{\psfig{figure=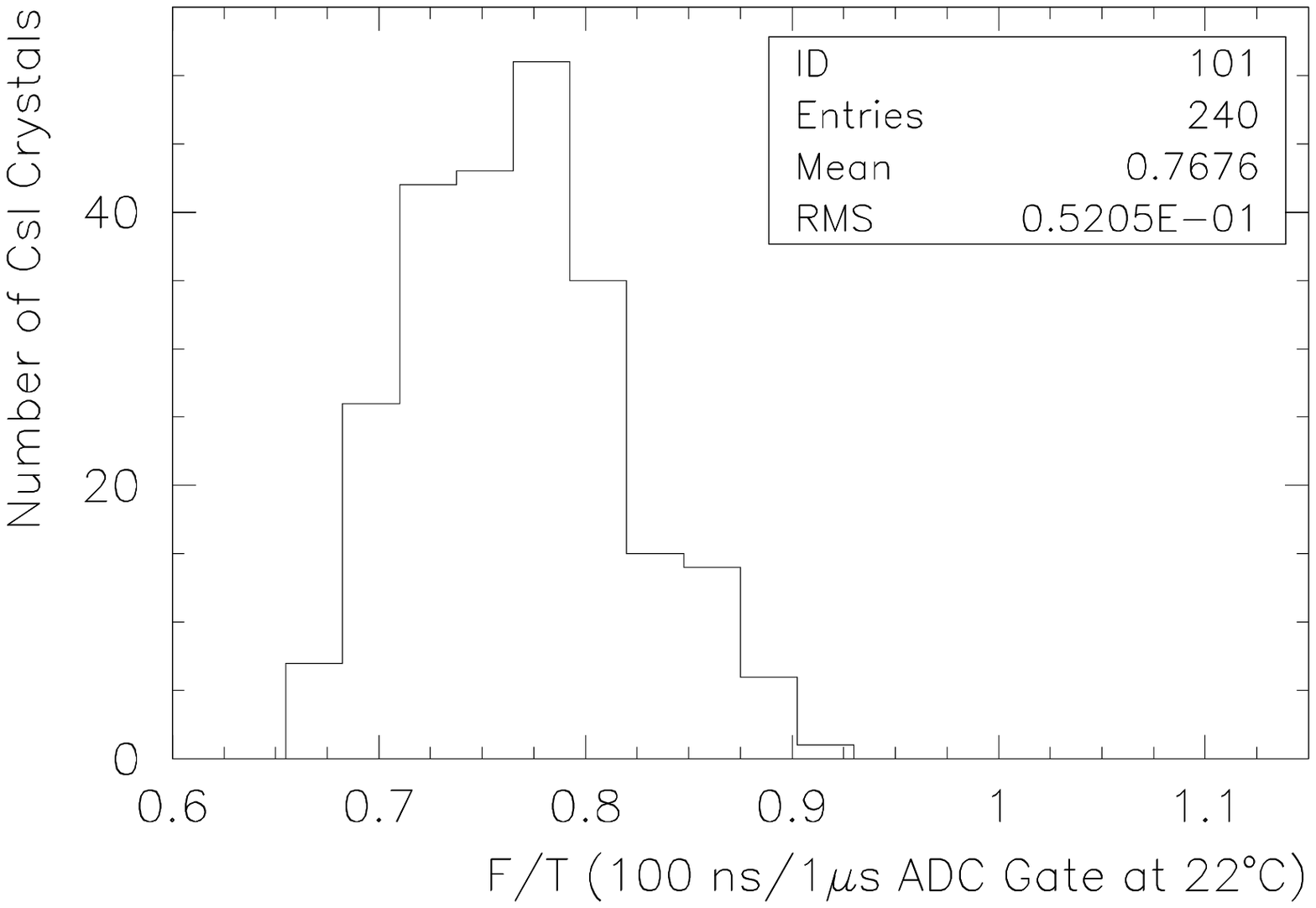,height=22cm}}
\vglue -8.50cm
\centerline{FIGURE 3}
\vspace*{\stretch{2}}
\clearpage

\vspace*{\stretch{1}}
\centerline{\psfig{figure=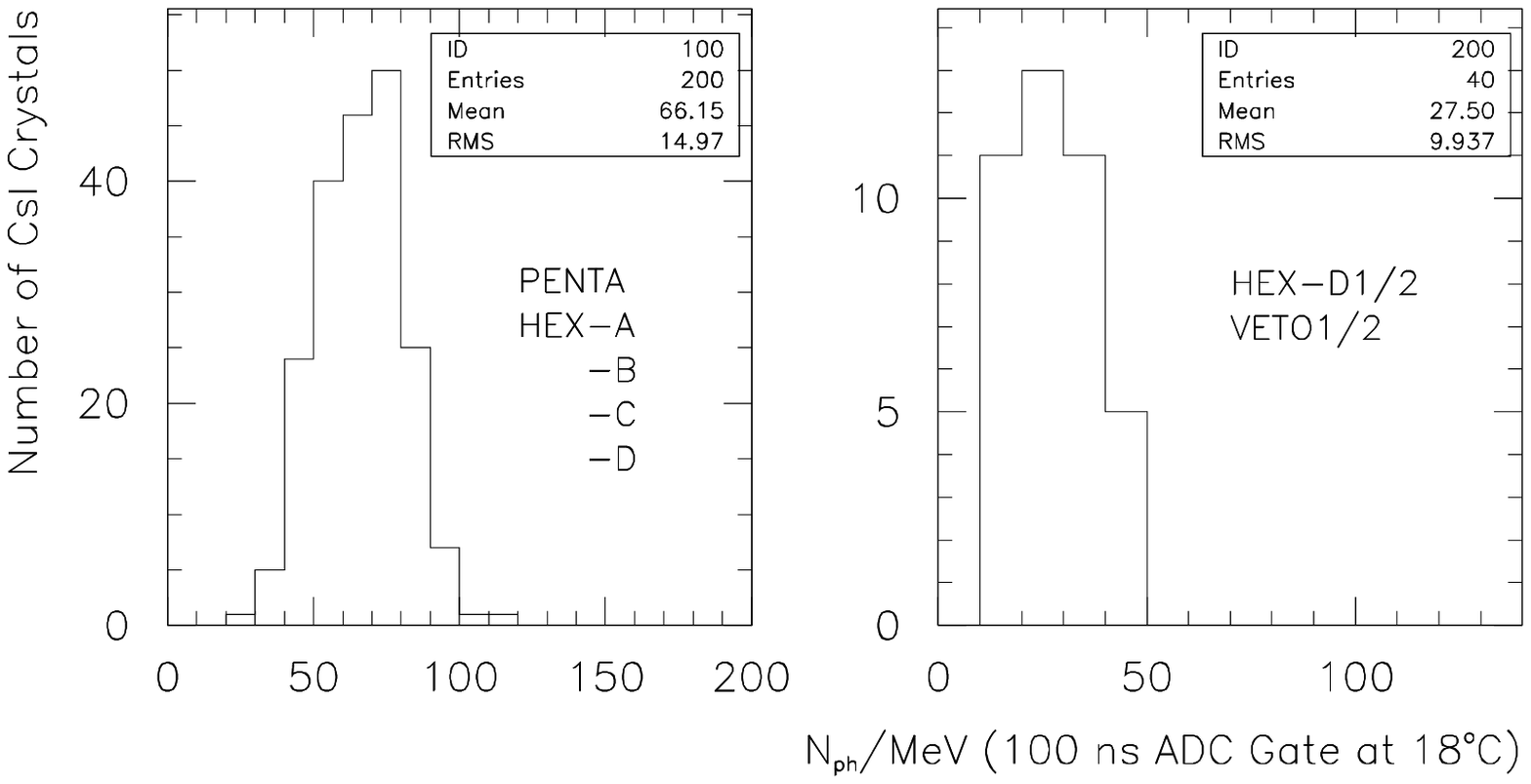,height=20cm}}
\vglue -8.00cm
\centerline{FIGURE 4}
\vspace*{\stretch{2}}
\clearpage

\vspace*{\stretch{1}}
\centerline{\psfig{figure=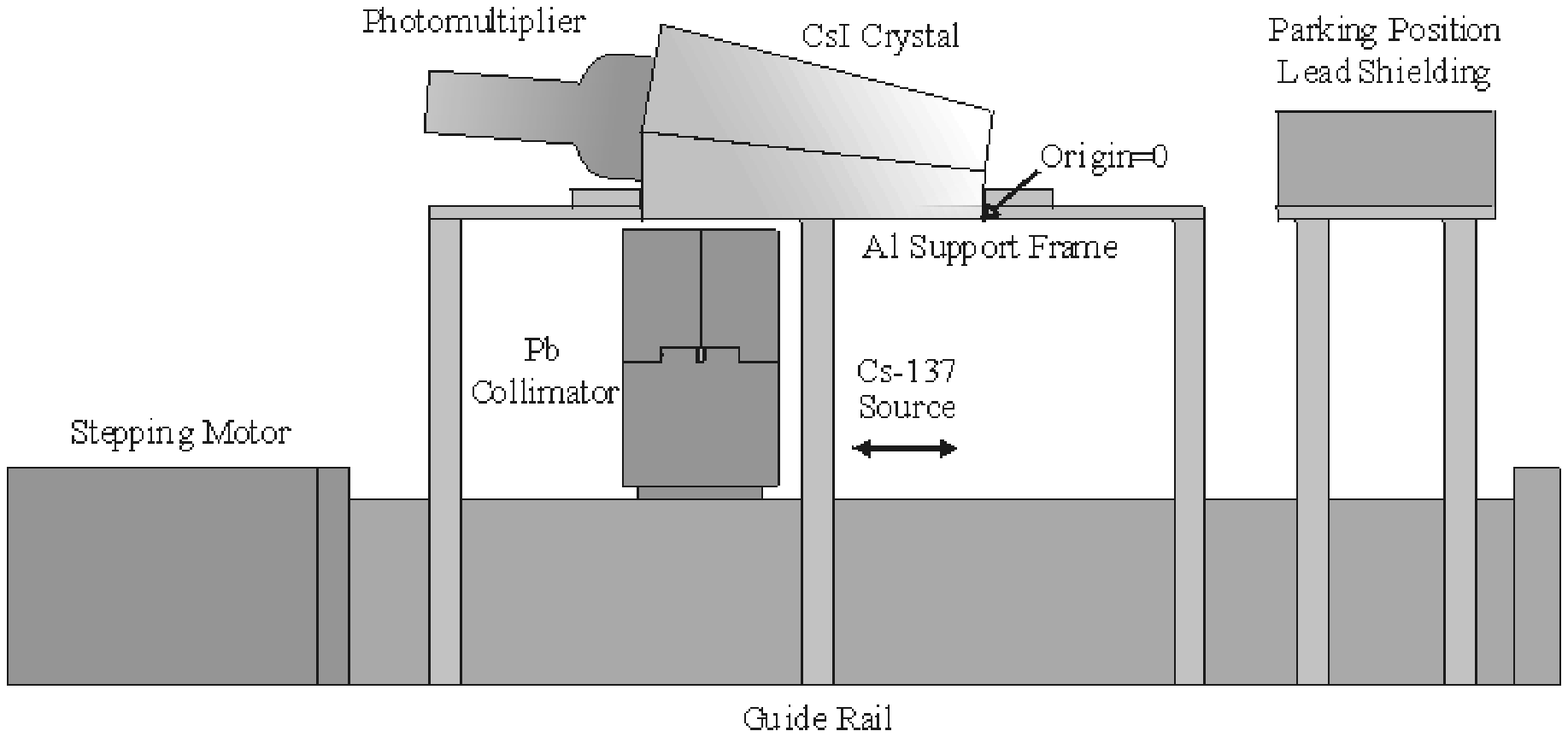,width=15cm}}
\bigskip\bigskip\bigskip\bigskip
\centerline{FIGURE 5}
\vspace*{\stretch{2}}
\clearpage

\vspace*{\stretch{1}}
\centerline{\psfig{figure=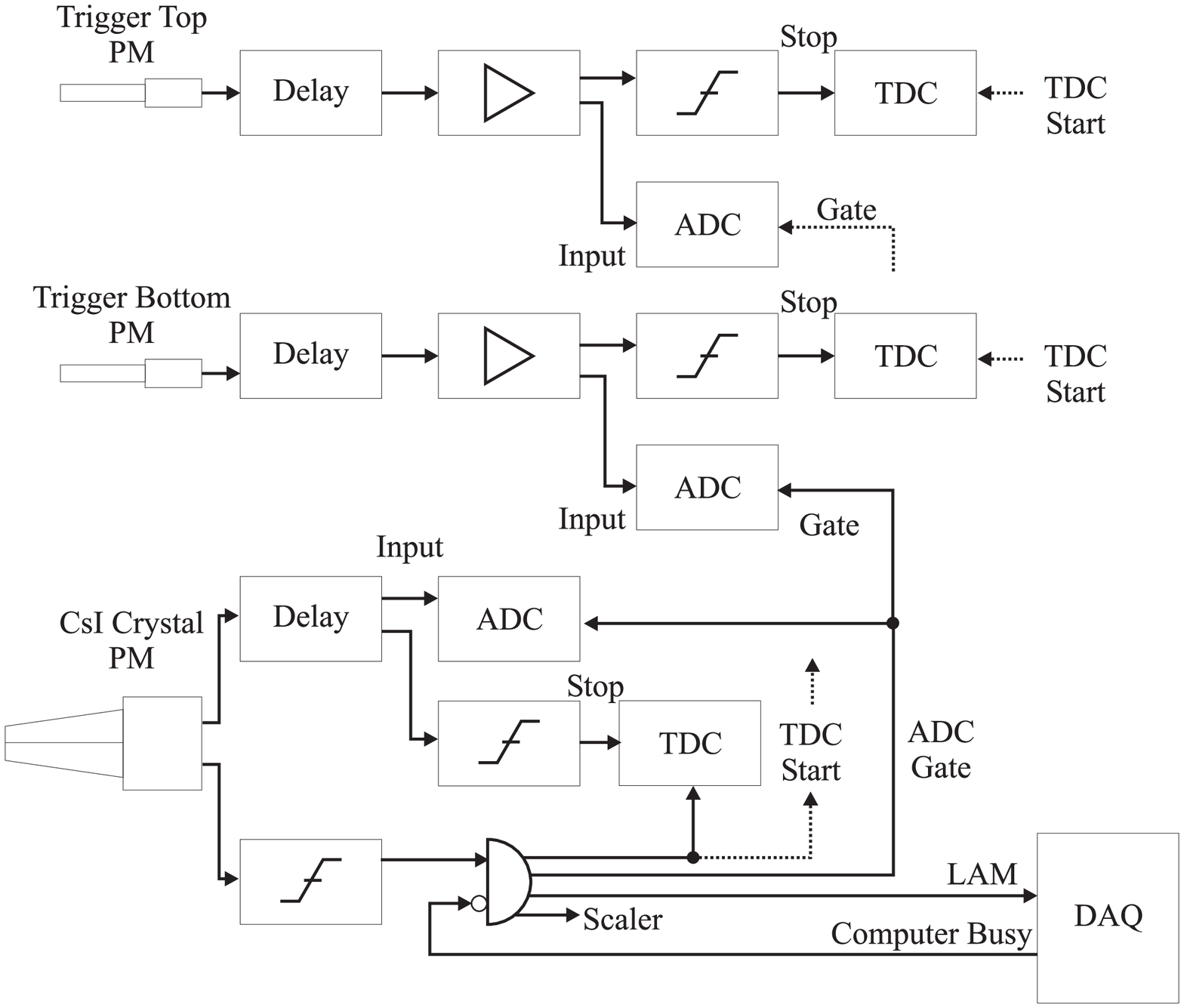,width=15.0cm}}
\bigskip\bigskip
\centerline{FIGURE 6}
\vspace*{\stretch{2}}
\clearpage

\vspace*{\stretch{1}}
\centerline{\psfig{figure=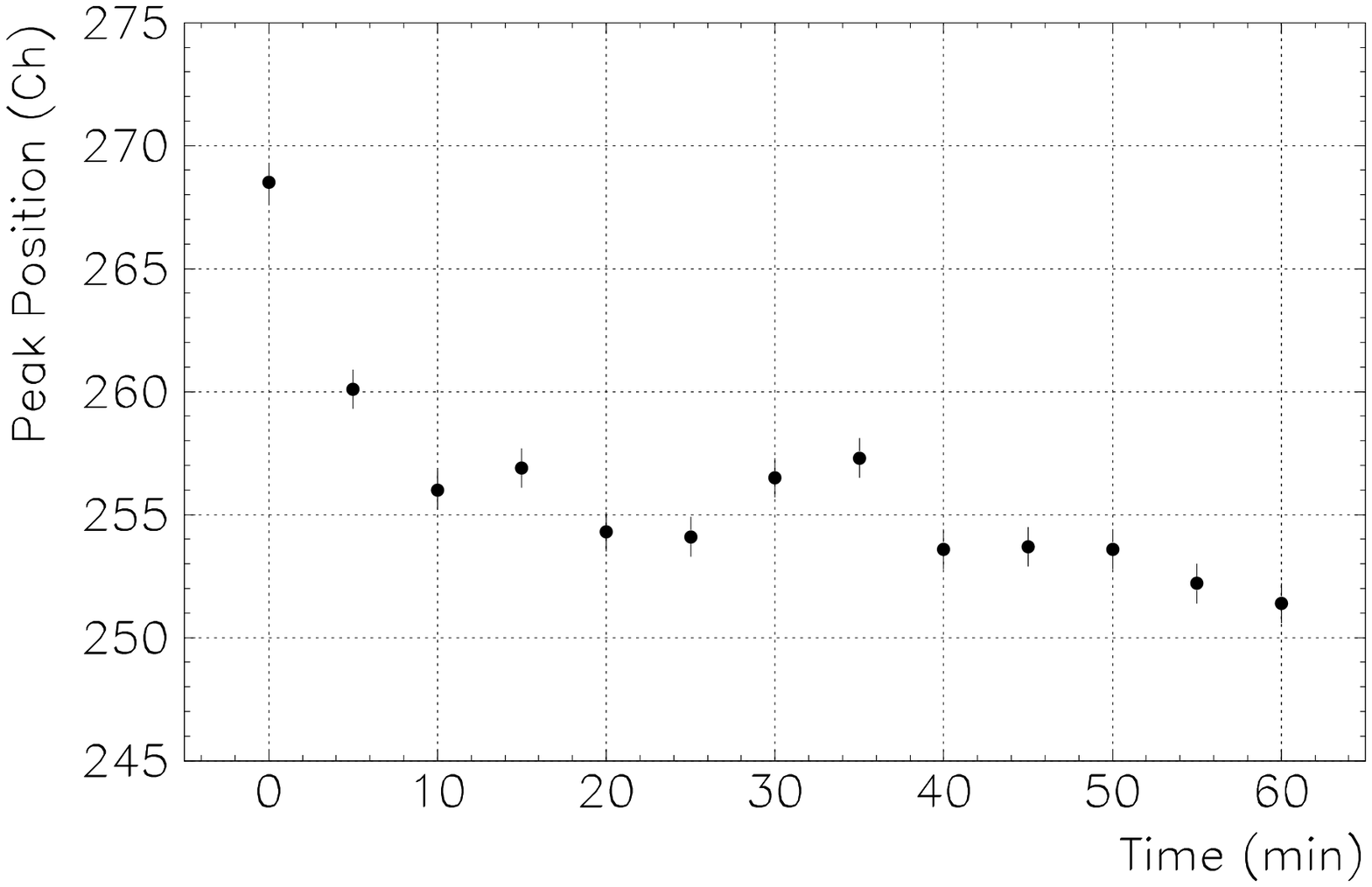,height=20.0cm}}
\vglue -8.7cm
\centerline{FIGURE 7}
\vspace*{\stretch{2}}
\clearpage

\vspace*{\stretch{1}}
\centerline{\psfig{figure=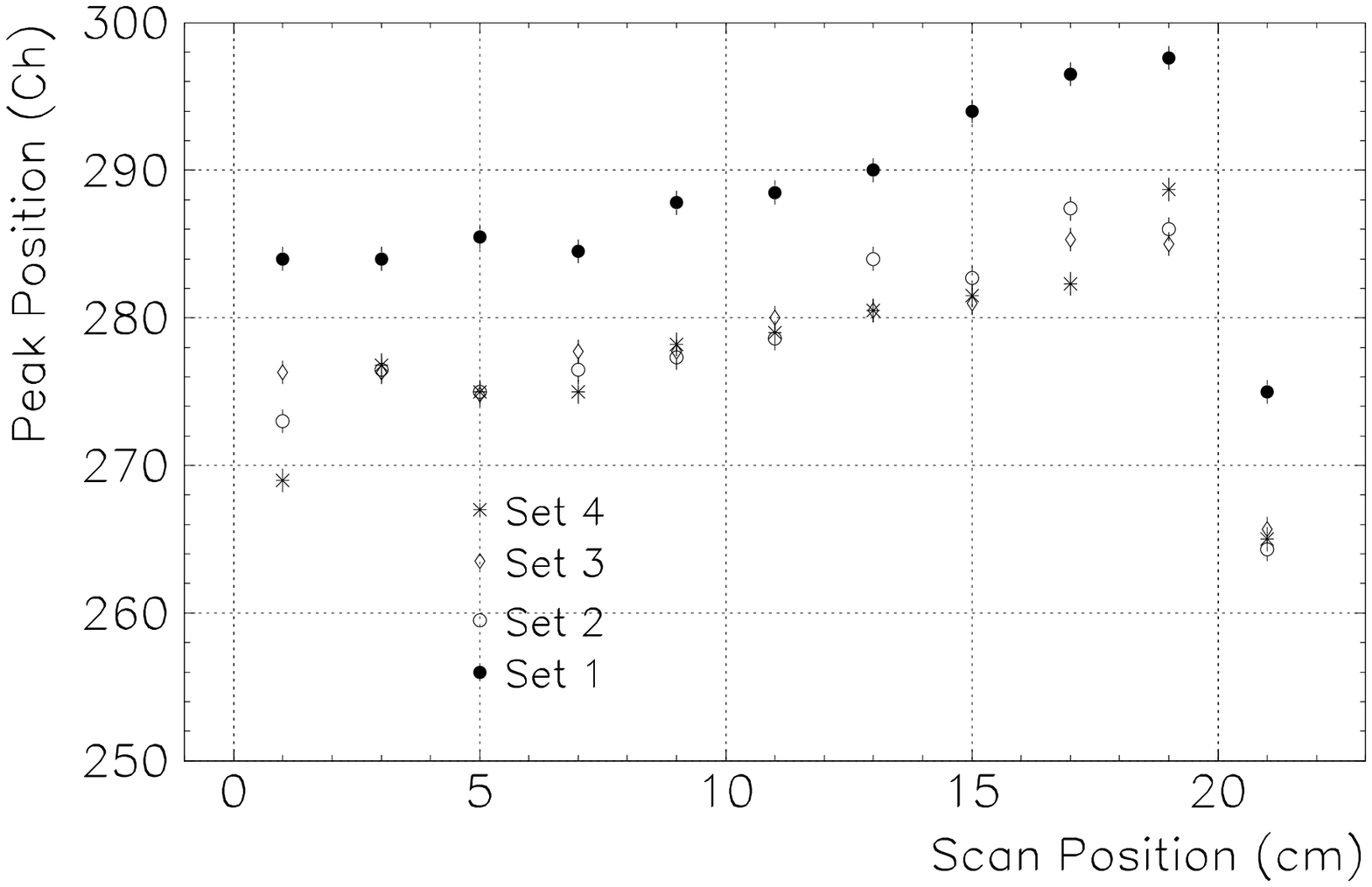,height=20.0cm}}
\vglue -8.5cm
\centerline{FIGURE 8}
\vspace*{\stretch{2}}
\clearpage

\vspace*{\stretch{1}}
\centerline{\psfig{figure=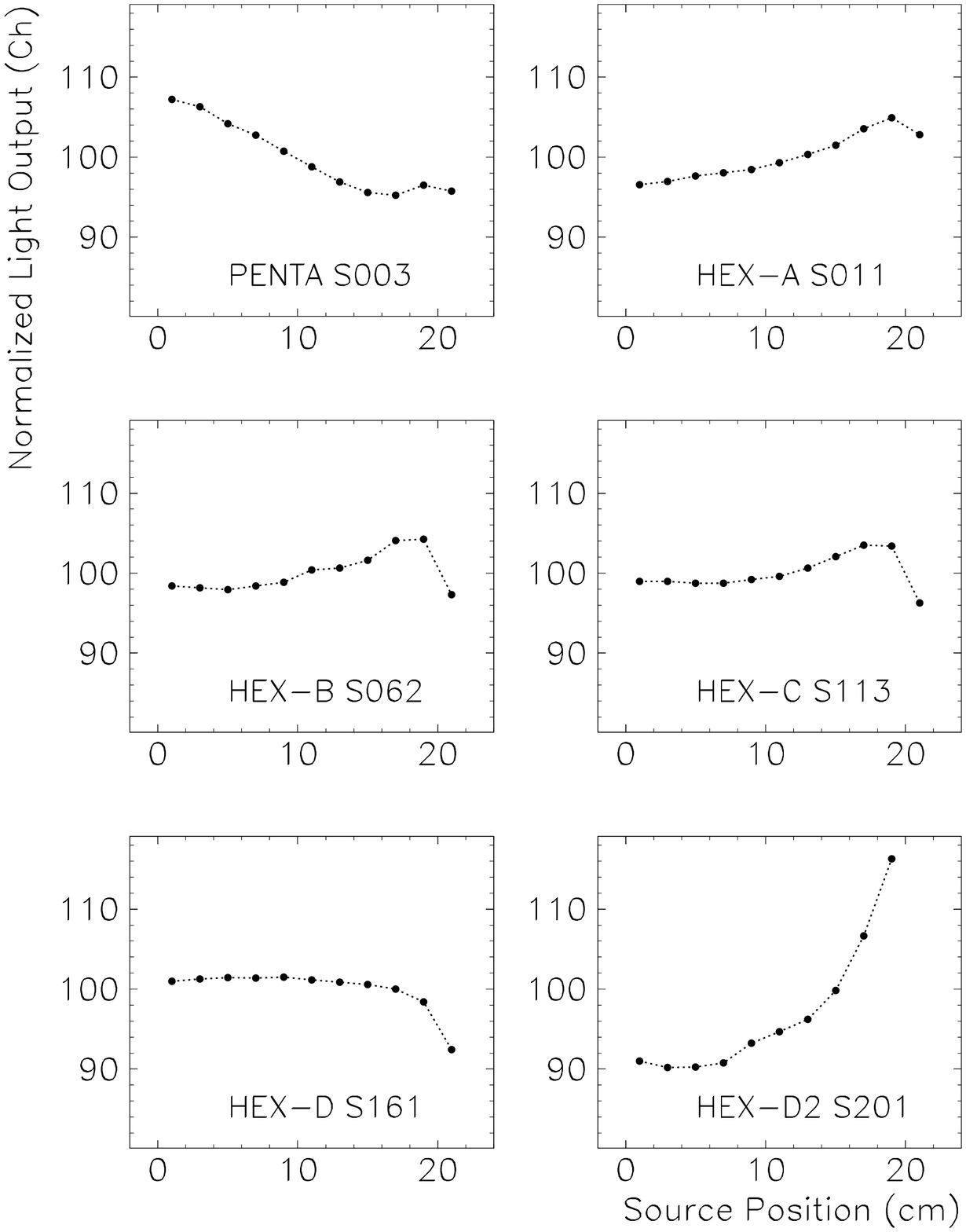,height=20cm}}
\vglue 0.5cm
\centerline{FIGURE 9}
\vspace*{\stretch{2}}
\clearpage

\vspace*{\stretch{1}}
\centerline{\psfig{figure=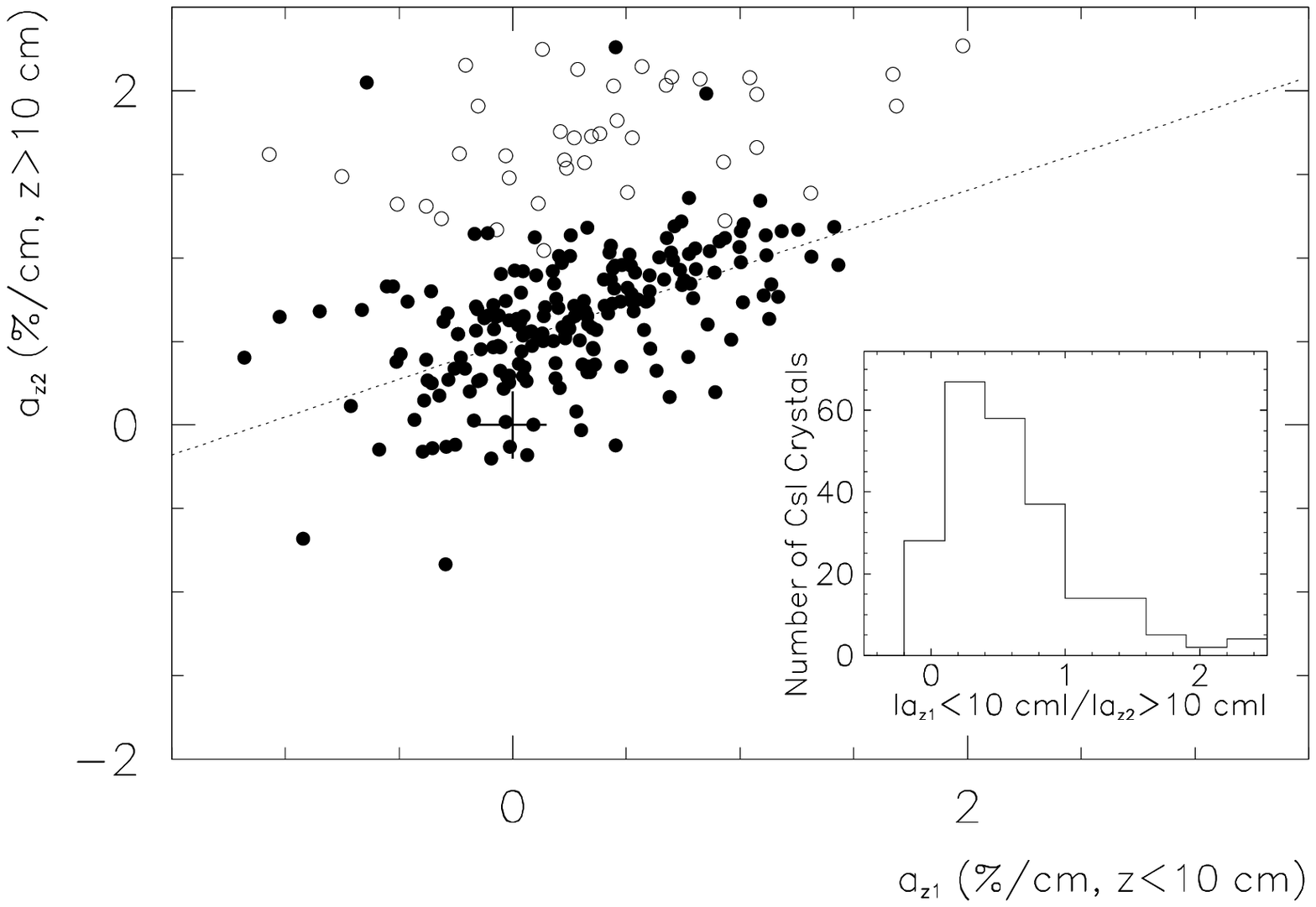,height=22cm}}
\vglue -8.7cm
\centerline{FIGURE 10}
\vspace*{\stretch{2}}
\clearpage

\vspace*{\stretch{1}}
\centerline{\psfig{figure=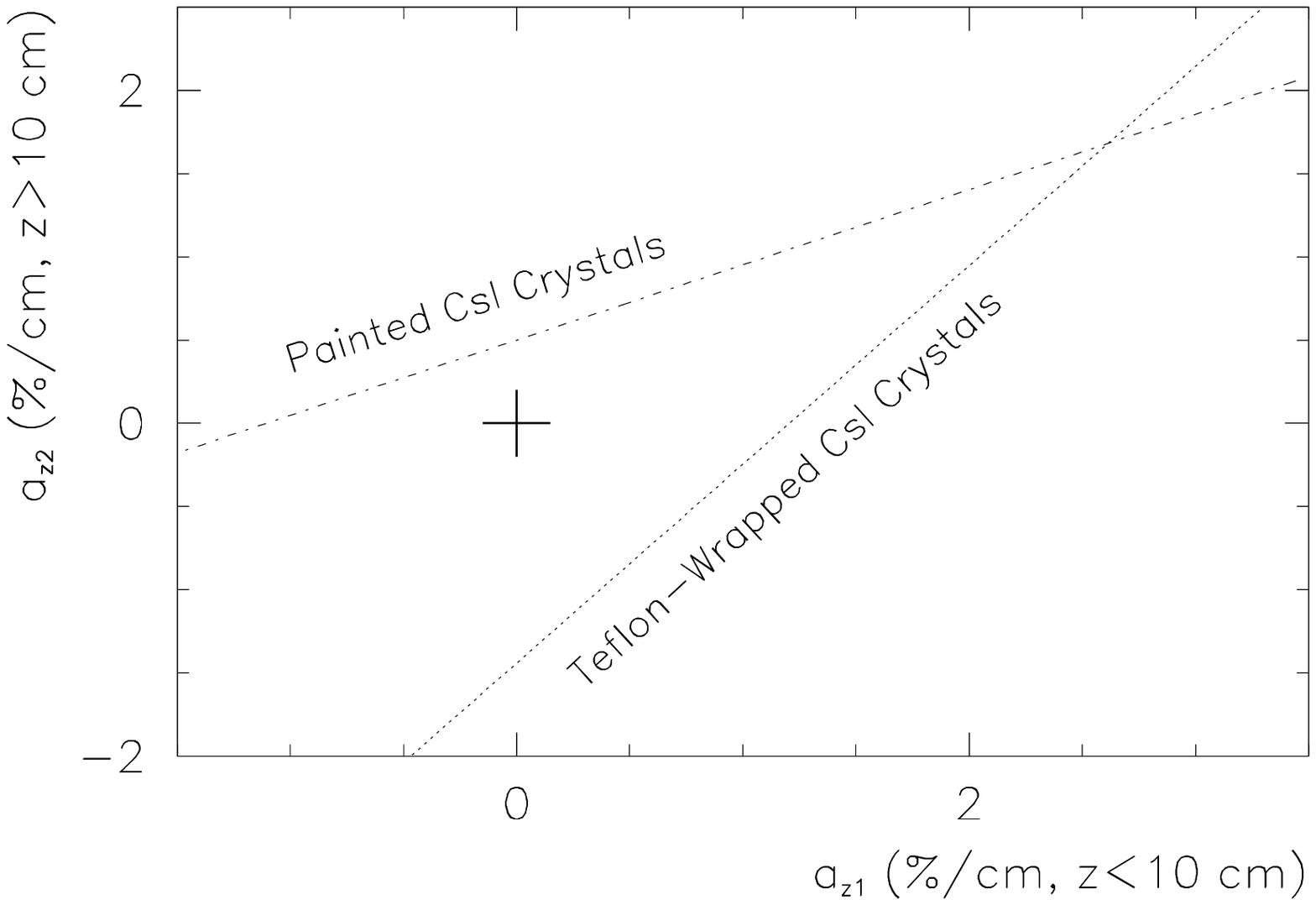,height=22cm}}
\vglue -8.7cm
\centerline{FIGURE 11}
\vspace*{\stretch{2}}
\clearpage

\vspace*{\stretch{1}}
\vglue -3cm 
\centerline{\psfig{figure=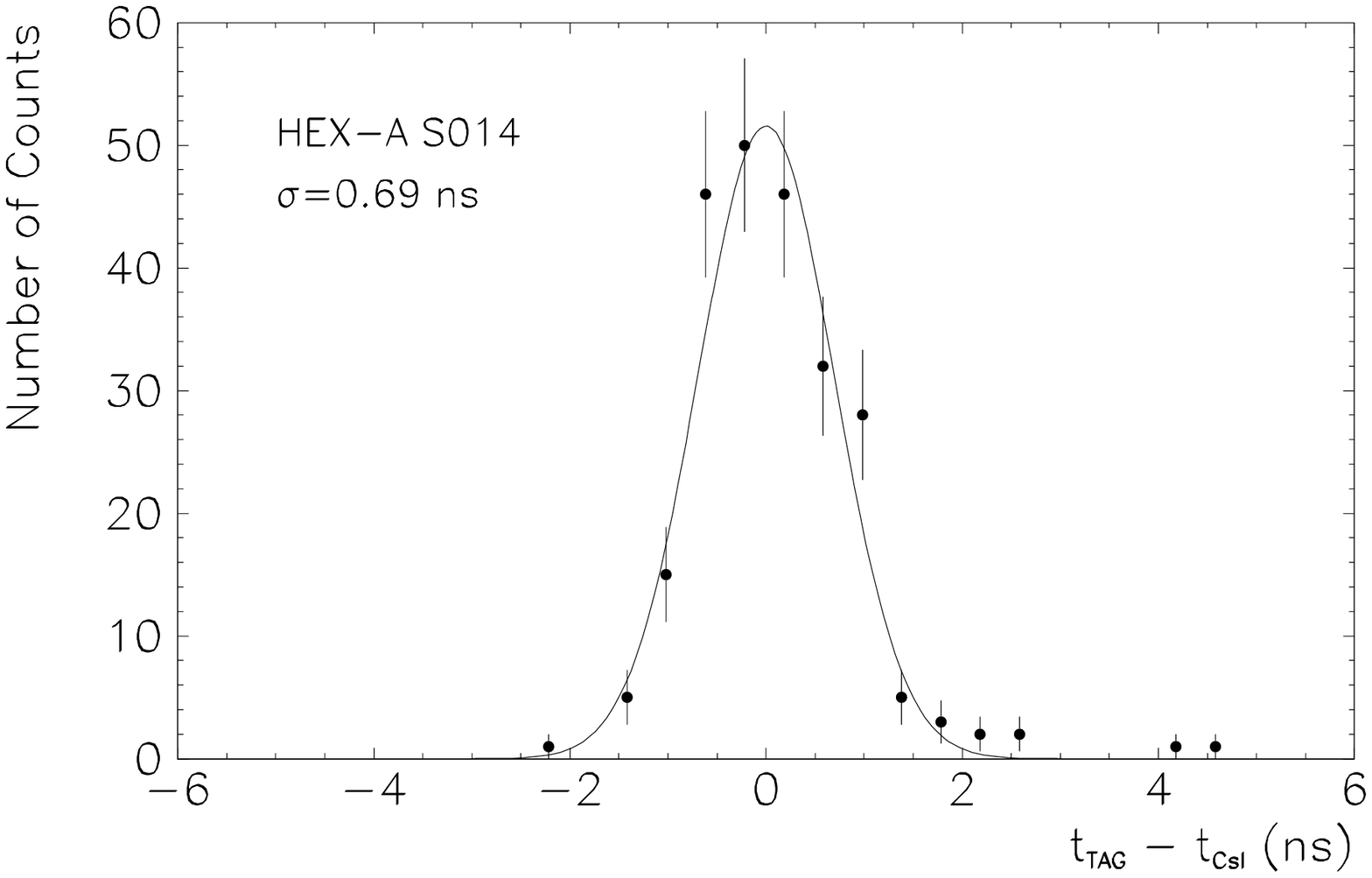,height=20cm}}
\vglue -9.5cm
\centerline{\psfig{figure=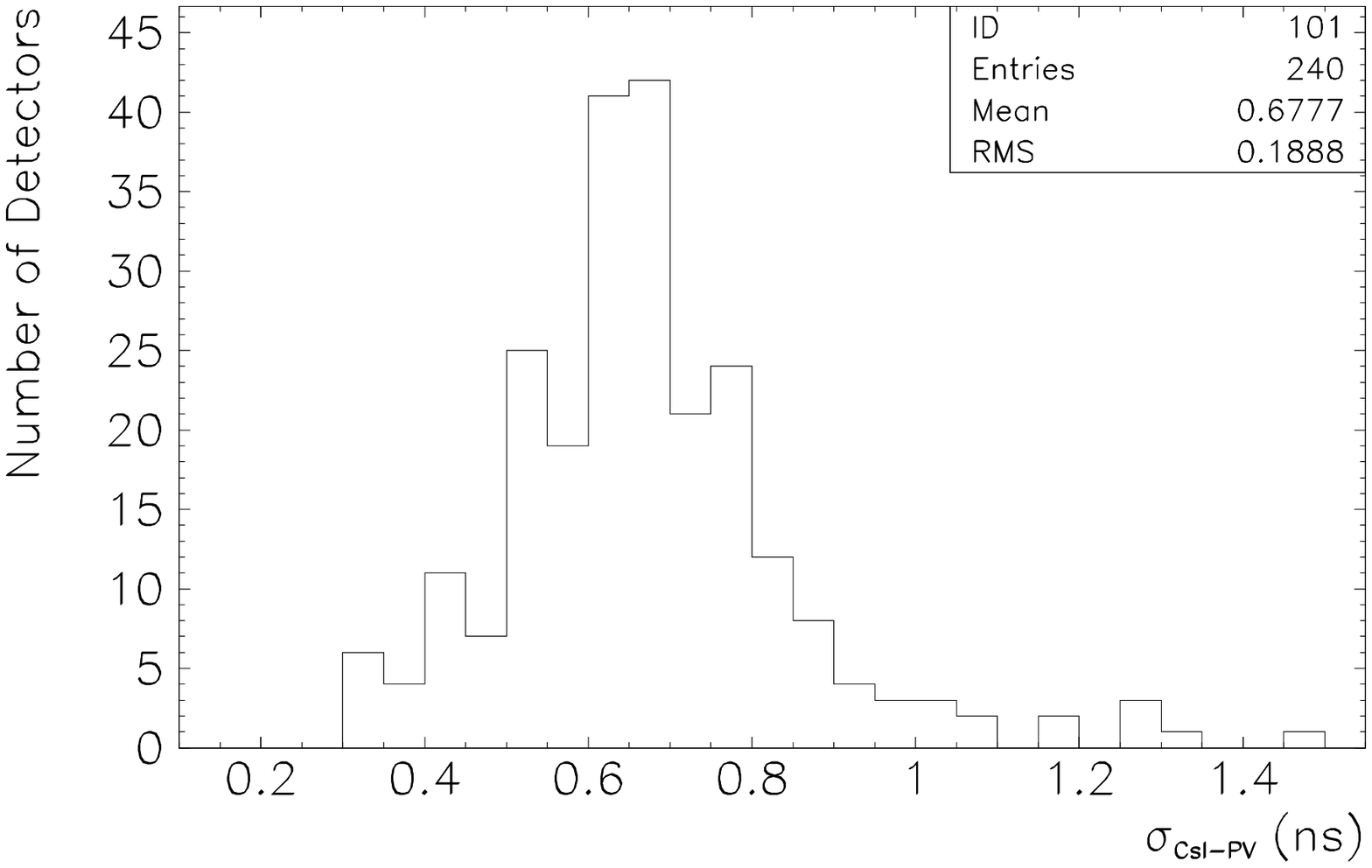,height=20cm}}
\vglue -9.0cm
\centerline{FIGURE 12}
\vspace*{\stretch{2}}
\clearpage

\vspace*{\stretch{1}}
\centerline{\psfig{figure=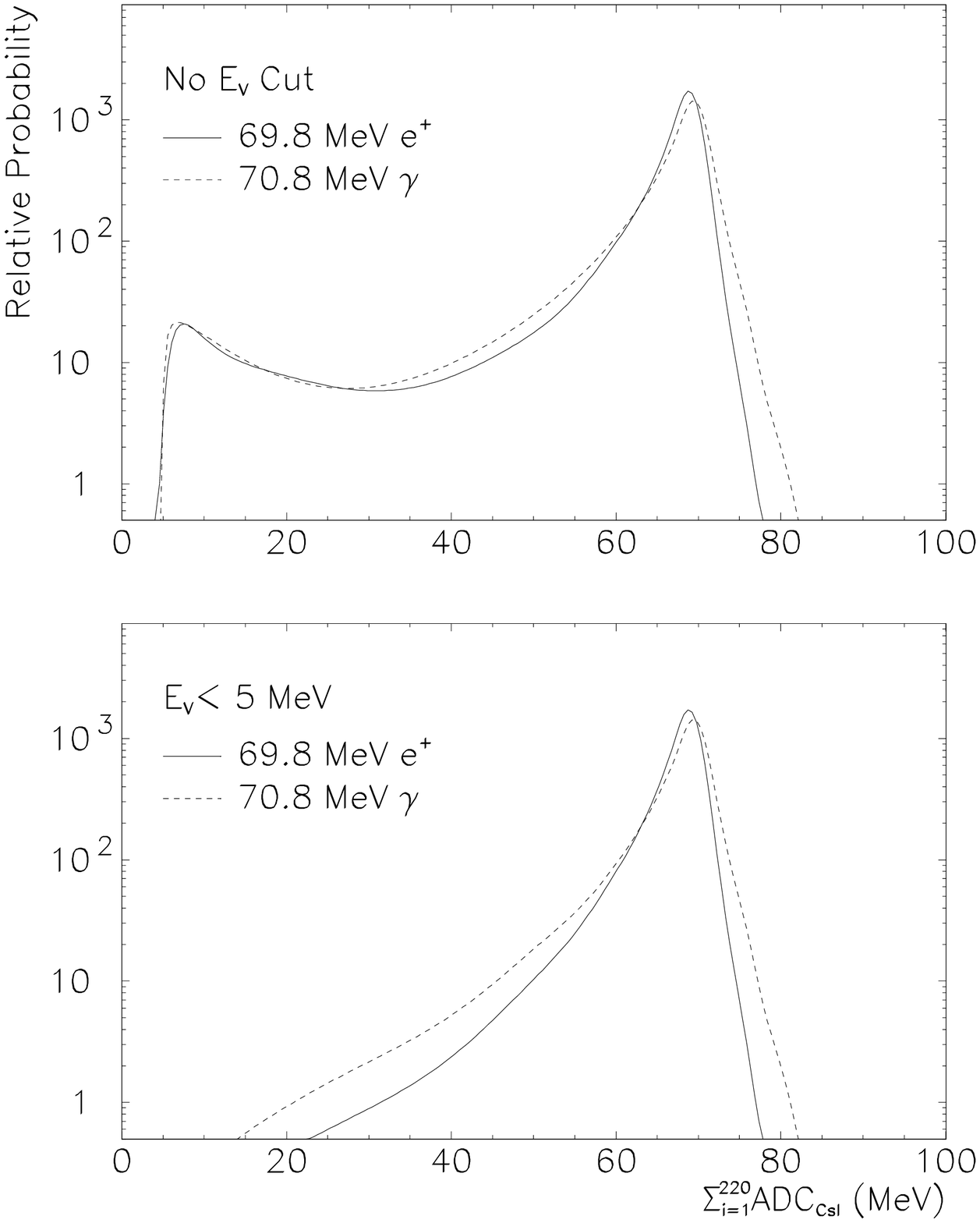,height=20cm}}
\vglue 0.5cm
\centerline{FIGURE 13}
\vspace*{\stretch{2}}
\clearpage

\vspace*{\stretch{1}}
\centerline{\psfig{figure=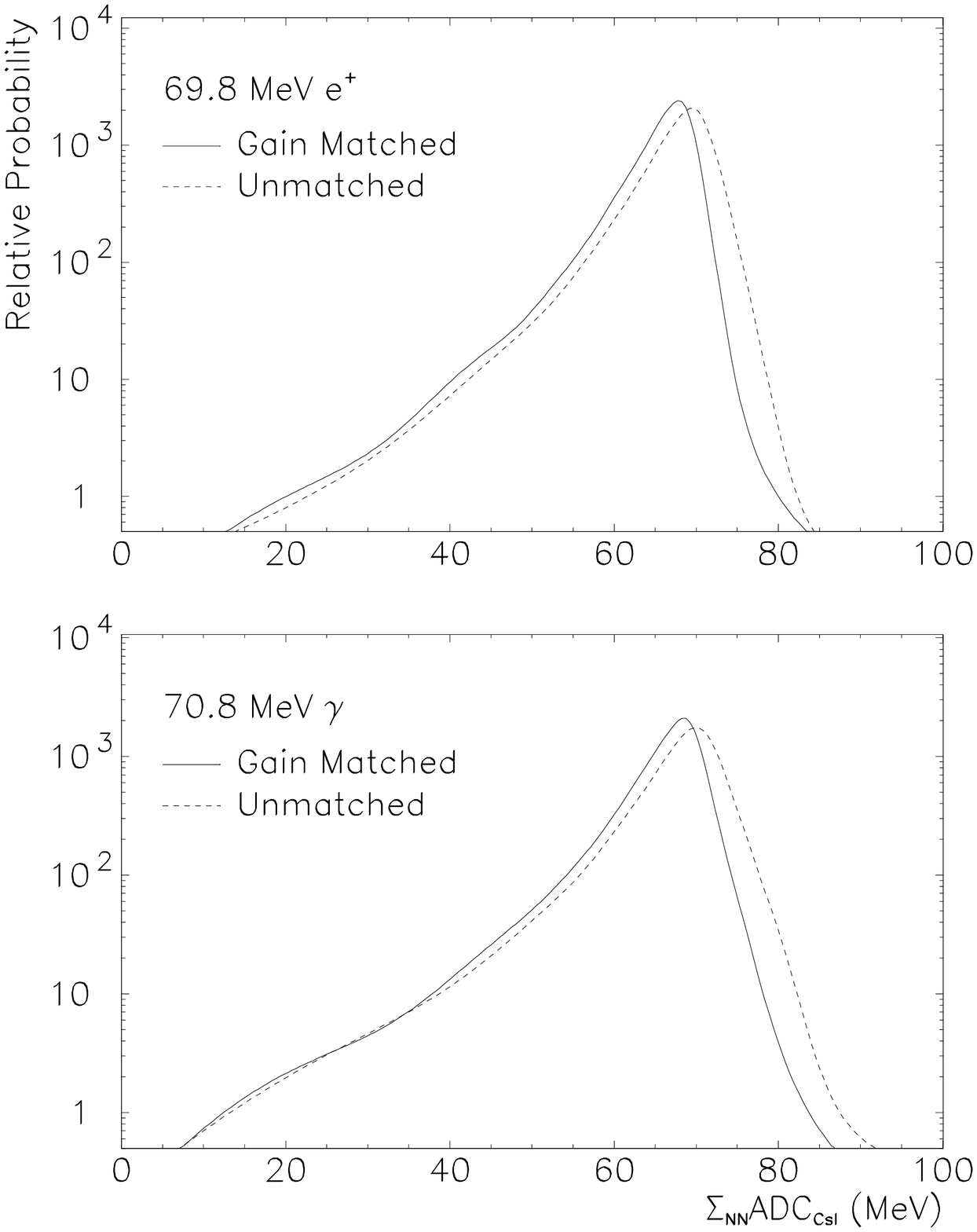,height=20cm}}
\vglue 0.5cm
\centerline{FIGURE 14}
\vspace*{\stretch{2}}
\clearpage

\vspace*{\stretch{1}}
\centerline{\psfig{figure=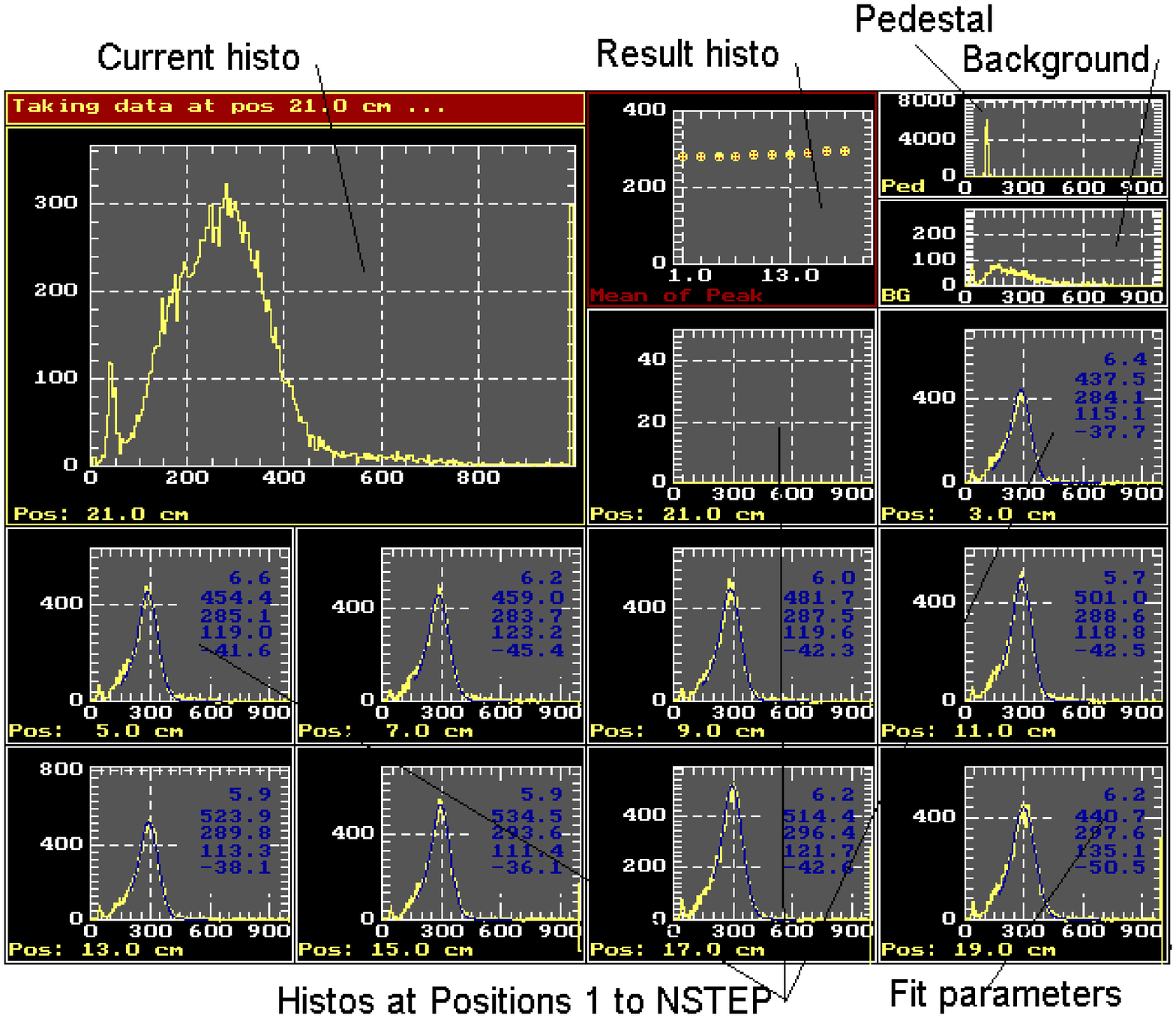,width=18.0cm}}
\vglue 1.5cm
\centerline{FIGURE 15}
\vspace*{\stretch{2}}
\clearpage

\end{document}